%% file: VoiceConversion.tex
\documentclass[a4paper]{article}

\usepackage{INTERSPEECH2019}

\title{Hierarchical Sequence to Sequence Voice Conversion with Limited Data}
\name{Praveen Narayanan, Punarjay Chakravarty, Francois Charette, Gint Puskorius}
\address{
  Ford Greenfield Labs, 
  Palo Alto, CA 
  }
\email{\{pnaray11,pchakra5,fcharett,gpuskori\}@ford.com}

\usepackage{multirow}

\begin{document}

\maketitle
\begin{abstract}
We present a voice conversion solution using recurrent sequence to sequence modeling for DNNs. Our solution takes advantage of recent advances in attention based modeling in the fields of Neural Machine Translation (NMT), Text-to-Speech (TTS) and Automatic Speech Recognition (ASR). The problem consists of converting between voices in a parallel setting when {\it $<$source,target$>$} audio pairs are available. Our seq2seq architecture makes use of a hierarchical encoder to summarize input audio frames. On the decoder side, we use an attention based architecture used in recent TTS works. Since there is a dearth of large multispeaker voice conversion databases needed for training DNNs, we resort to training the network with a large single speaker dataset as an autoencoder. This is then adapted for the smaller multispeaker voice conversion datasets available for voice conversion. In contrast with other voice conversion works that use $F_0$, duration and linguistic features, our system uses mel spectrograms as the audio representation. Output mel frames are converted back to audio using a wavenet vocoder. 


\end{abstract}

\section{Introduction}
Recently, sequence to sequence models have been adapted with great success in producing realistic sounding speech in TTS systems \cite{Tacotron, Tacotron2, DeepVoice1, DeepVoice2, DeepVoice3}. 
Likewise, it has been demonstrated that ASR can be handled excellently by seq2seq architectures. In TTS, the system takes in a text or phoneme sequence and outputs a speech representation as output. On the other hand, in ASR, one feeds in an audio representation, and the system performs the task of classifying audio into text or phoneme. In voice conversion, both the input and output sequences are audio representations. The problem is related to both ASR and TTS in that like ASR, the DNN must learn to summarize input audio frames into a hidden context, and like in TTS, it must decode audio frames from the latent context in a temporal, attentive fashion. 

In voice conversion, we seek to convert a speech utterance from a source speaker A to make it sound like an utterance from a target speaker B. There are two pertinent scenarios, the first of which is when both the source and target speakers are uttering the same text (the 'parallel' case), and the second is when the utterances don't match (the 'non-parallel' case). We focus on parallel voice conversion in this work with DNNs. While the larger goal of this work is to address the more important problem of non-parallel voice conversion (producing parallel datasets for conversion is not easy), we start with the arguably simpler task of demonstrating how we can achieve this in the parallel scenario using seq2seq models. 

While we can go about the voice conversion task by first performing ASR on the source voice, and then sending the text obtained to a TTS engine, our approach leads to an end-to-end solution wherein one doesn't have to train an ASR and TTS engine separately.
Our approach has a simpler processing pipeline as it only needs audio transcripts (with no accompanying text or need for segmentation), and can be converted directly to target representations. A notable aspect of this work is that it gets around the problem of limited parallel data for voice conversion by pretraining a much larger, single speaker TTS corpus as an autoencoder and then performing transfer learning on the available, diminutive voice conversion datasets. Without this adaptation technique it becomes difficult to effectively carry out voice conversion without having access to larger, expensive to obtain parallel datasets. 

We train the system using Maximum Likelihood to minimize the L1 error between the generated and target mel spectrograms.

\begin{figure}
    \centering
    \includegraphics[width=0.99\linewidth]{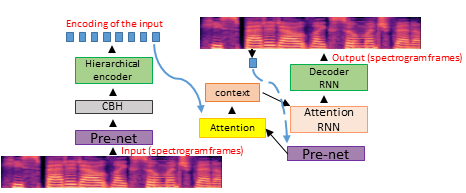}
    \caption{System Diagram: Our Attention based Encoder-Decoder architecture for Voice Conversion takes in a mel-spectrogram for the source speaker and outputs the mel-spectrogram for the target speaker.}
    \label{fig:system_diag}
\end{figure}

\section{Related Work}


The traditional pipeline for parallel voice conversion is through use of Gaussian Mixture Models (GMMs) \cite{Spectral-VC-Kain, Toda, Stylianou-GMM} or Deep Neural Networks (DNNs) \cite{DNN-Desai1, DNN-Desai2, Sun-BiLSTM, Sun-phonetic-posteriograms-vc}. After first aligning source and target features using Dynamic Time Warping (DTW)\cite{DTW-book}, the model is trained so that it learns to produce the target given the source features for each frame. A disadvantage of these methods is that they need aligned, parallel data. Moreover, conversions performed on spectral features disregard dependencies on other controlling factors such as prosody and fundamental frequency, duration and rhythm. Furthermore, transforming features on a frame basis disregards temporal context dependencies. The dependence on fundamental frequency is often handled by performing a linear transformation in the logarithmic domain. A good review article of the topic is found in \cite{Mohammadi-review}. 

Non-negative Matrix Factorization (NMF) \cite{li2019robust}, traditionally used for sound source separation and speech enhancement, has also been used for VC \cite{takashima2013exemplar}. NMF factorizes a matrix into two non-negative factors, the basis or dictionary matrix and the activation matrix. In the case of VC with parallel training data, dictionaries for speaker 1 and speaker 2 are first constructed separately. Subsequently, given test source data (speaker 1), the previously learnt dictionary for speaker 1 is used to factorize the source voice into a set of source activations, or contributions of speaker 1 dictionary to speaker 1 utterance. The same is done for speaker 2. The activations for the source speaker utterance are then combined with the dictionary atoms of the target speaker utterance to transform speaker 1 utterance into speaker 2. NMF based methods, like GMMs, also require alignment of parallel voice samples using Dynamic Programming, and other pre-processing steps like the Short-term-Fourier-Transform (STFT). 

Recent sequence to sequence modeling approaches for voice conversion have largely been inspired by advances in seq2seq practice in NMT, TTS and ASR, in that they involve an encoder-decoder model as the underlying machinery. It is often advantageous to classify the input waveform into text or phoneme, and use that information to inform the decoder model of the content that the input audio representation embodies \cite{SCENT-model, SCENT-model2}. Our work is most similar to \cite{SCENT-model}, and we compare and contrast salient aspects of both models. In both models, the overall architecture is a seq2seq model inspired by the ASR work \cite{LAS} (with a hierarchical encoder stack), and the TTS work Tacotron \cite{Tacotron} and derivatives. However, in \cite{SCENT-model}, the encoder outputs are augmented by features extracted with an ASR model, while our approach comprises end-to-end neural networks without need for labeled data. There are also several ancillary components such as the use of additional losses and postprocessing networks. Also, we use a Convolutional filter bank, Highway network to extract 'context' as a prelude to processing them in the multilayer hierarchy of encoder RNNs. Nevertheless, we wish to emphasize that a substantive difference between the two works is the training philosophy, in how the data limitation problem is handled. We elaborate on this in the following paragraphs with additional examples from the literature. 

Pertinent to our discussion are seq2seq modeling works \cite{Tanaka-seq2seq, convS2S-vc}. In these works, additional loss terms are introduced to encourage the model to learn alignment and to preserve linguistic context. Alignment is maintained by noting that the attention curve is predominantly diagonal (in the voice conversion problem) between source and target, and including in the loss function a diagonal penalty matrix - a term referred to as guided attention in the TTS work \cite{Tachibana-Uenoyama}. An additional consideration is to prevent the decoder from 'losing' linguistic context, as would arise when it simply learns to reconstruct the output of the target. This was addressed by using additional neural networks that ensure that the hidden representation produced by the encoder (similar reasoning applies to the decoder) was capable of reconstructing the input, and thereby retained context information. These manifest as additional loss terms - we also glean a similarity to cycle consistency losses \cite{CycleGAN} - that they call 'context preservation losses'. Also noteworthy is that these approaches use non-recurrent architectures for their seq2seq modeling.

We suspect that the problems that motivated the design of these additional losses have their provenance in the diminutive size of the training corpus (CMU Arctic was used, with $1132$ utterances per speaker) which is hardly sufficient to learn a diverse respresentation with good generalization capabilities. In our work, we arrive at a slightly different way to overcome the data limitation problem as compared with these works, which do so by augmenting data with ASR training \cite{SCENT-model, SCENT-model2}, and by introducing additional losses \cite{Tanaka-seq2seq, convS2S-vc}.  Our solution makes use of transfer learning by first pretraining with a large, single speaker corpus, and then adapting to the smaller, pertinent corpus (CMU Arctic) in question.

Developments in the generative modeling (primarily, Variational Autoencoders \cite{VAE} and Generative Adversarial Networks \cite{GAN}) front have led to their use in voice conversion problems. In \cite{Kaneko-LearnedSimilarityMetric}, a learned similarity metric obtained through a GAN discriminator is used to correct oversmoothed speech that results from maximum likelihood training, which imposes a particular form for the loss function (usually the MSE). A conditional VAEGAN \cite{VAEGAN} setup is used in \cite{VAW-GAN} to implement voice conversion, with conditioning on speakers, together with a Wasserstein GAN discriminator \cite{WGAN} to fix the blurriness issue associated with VAEs. Moreover, an important apparatus that is of use in training non-parallel voice setups consists of Cycle Consistency Losses from the famous CycleGAN \cite{CycleGAN} work for images. This forms a building block in the papers \cite{Kaneko-CycleGAN} and \cite{Kameoka-StarGAN}. 

A natural extension to our work is to explore a generative solution to Voice Conversion as in some of the works above, in order to apply our architectural components to non-parallel setups. 

Our work is influenced by recent TTS works involving transfer learning and speaker adaptation. The recently published work \cite{Wavenet-Few-shot-adaptation} demonstrates a methodology to use adapt a trained network for new speakers with a wavenet. Likewise, in \cite{Tacotron-transfer-learning}, a speaker embedding is extracted using a discriminative network for unseen, new speakers which is then used to condition a TTS pipeline similar to Tacotron. This philosophy is also used in \cite{VoiceCloningFewSamples} where schemes are used to learn speaker embeddings extracted separately or trained as part of the model during adaptation. In all these contexts, it is emphasized that the onus is on adapting to small, limited data corpuses, thereby circumventing the need to obtain large datasets to train these models from scratch. In our work, we use the same idea to get around the problem of not having enough data to train in the voice conversion dataset under consideration. However, in our work, instead of producing new speaker embeddings, we retrain the model for each new $<source,target>$ pair, a process that is rapid owing to the small size of the corpus. 

An interesting alternative to using recurrent (or autoregressive) seq2seq modeling for TTS or VC is to use differential memory as a way to store speech related information. In the VoiceLoop architecture \cite{Taigman-VoiceLoop, Nachmani-VoiceLoop1, Nachmani-VoiceLoop2}, the input is transformed with a shallow fully connected network into a context, with attention being used to compare with the memory buffer. The memory buffer itself is updated by replacing its first element with a newly computed representation vector. With this approach, which also uses speaker embeddings, the network is able to adapt to new speakers with only a few samples, in addition to having a much reduced network complexity (only shallow fully connected layers are used).

\section{Architecture}
We use an attention based encoder-decoder network for our voice conversion task. The network architecture borrows heavily from recent developments in TTS \cite{Tacotron} and ASR \cite{LAS}. The system takes in an audio representation (mel-spectrogram) as input, and encodes it into a hidden representation in recurrent fashion. This hidden representation is then processed by an attention based decoder into output mel-spectrograms. In order to convert the mel frames back to audio, we employ a widely used wavenet vocoder implementation available online \cite{r9y9-wavenet}. In the Tacotron2 \cite{Tacotron2} work, it was demonstrated that using wavenet as a neural vocoder produced audio samples whose quality was superior to those from the Griffin-Lim procedure used in Tacotron \cite{Tacotron}.

A system diagram showing the various components of the model is shown in Figure \ref{fig:system_diag}.
We describe its components in the following subsections. 

\subsection{Encoder}
\subsubsection{Prenet}
The prenet is a bottleneck layer containing full connections with a ReLU nonlinearity and dropout \cite{Tacotron,Dropout}. The purpose of this layer is to enable the model to generalize to unseen input with dropout. Mechanisms to achieve this effect in sequence models are teacher forcing, scheduled sampling and professor forcing \cite{TeacherForcing, ScheduledSampling, ProfessorForcing}. Prenet processes vectors of 80 dimensions to yield output of the same size. A dropout ratio of $0.5$ is used. 

\subsubsection{CBH: Convolutional Banks and Highway layers}

\begin{figure}
    \centering
    \includegraphics[width=0.55\linewidth]{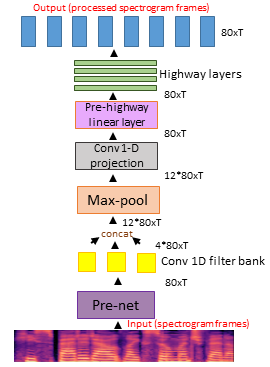}
    \caption{The Pre-net and the CBH layers that are used to process the input mel-spectrogram frames. Output tensor sizes at each step of processing are indicated by the side of the unit.}
    \label{fig:cbh}
\end{figure}

Originally proposed in the context of Neural Machine Translation \cite{Lee} and later used in \cite{Tacotron} where it was named CBHG (Convolutional  Banks, Highway and Gated Recurrent Units), this layer served as a processing mechanism to accumulate 'word' level context when the input is text. For our voice conversion task, the effect is similar, in that neighboring speech frames are filtered so as to abstract the equivalent phoneme level representation, mixed with speaker characteristics and prosodic content. Together with the hierarchical RNN encoder units  described later, we could view this assemblage as an implementation of CBHG. The Pre-net and CBH layers, along with the tensor output sizes at each step of the processing are shown in Figure ~\ref{fig:cbh} and described below.

\textbf{Convolutional Filter Banks}
A bank of 1-D convolutional filters of size $(1,3,5)$ are used to capture n-grams of varying width. The input sequence is convolved with each of these filters and the results from all filters are concatenated together. Each convolution filter preserves the original length of the sequence by padding it to extend original length by $w-1$, where the filter is of width $w$, followed by BatchNorm and ReLU operations. The filter maps obtained are then stacked in the channel dimension with $4$ output channels being produced for each convolution. This is followed by a max-pool operation with stride 1, which maintains the length of the sequence.
A 1-D convolutional projection operation then reduces the sequence length to the original size, followed by a final linear layer that also maintains representation length. 

\textbf{Highway Layers}
The Highway layer is like a Resnet block with a skip connection that is a shortcut for information flow that skips the intermediate layers, but with learnable weights to determine the extent of the information skip. We use 4 highway layers.


\subsubsection{Hierarchical Recurrent Encoder}
\begin{figure}
    \centering
    \includegraphics[width=0.55\linewidth]{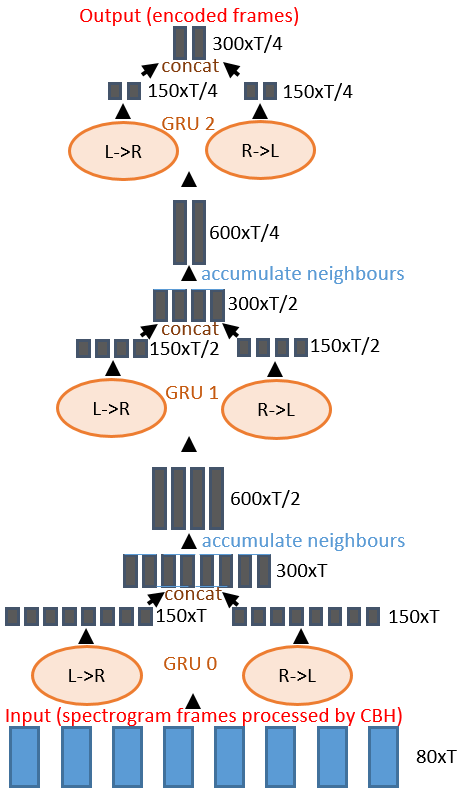}
    \caption{Hierarchical Bi-directional Recurrent Encoder with an indication of the tensor sizes at each step. The number of hidden units in each GRU is 150. Each pyramidal GRU unit (GRU 1 and 2) decreases the sequence length by 1/2. Left-right and right-left GRU units each output a 150xT matrix, that are concatenated to give a 300xT matrix, with T as input sequence length.}
    \label{fig:encoder}
\end{figure}

We design our encoder as a stack of bidirectional layers, reducing the sequence length by a factor of $2$ as the data flows up the stack (Figure \ref{fig:encoder}). This construction was first proposed in \cite{LAS} in the context of speech recognition with DNNs. The encoder's task is to summarize audio input to an intermediate hidden representation embodying linguistic content, akin to text. However (and this might be argued as a desirable attribute of DNN processing), we make no attempt to disentangle content (text) and voice characteristics (style) in this case. We assume that the DNN automatically learns to disentangle content and style as part of the training process, and that during the decoding process, the first speaker's voice characteristics are discarded and the second speaker's voice is injected into the content. 

The reasoning behind using a hierarchical reduction of timesteps is that speech frames are highly inflated, redundant descriptors of linguistic content mixed with speaker and duration information. A single phoneme could thus span several ($\sim$10) frames. It therefore makes sense to reduce or cluster the speech frames so as to contain more relevant information. By reducing the number of timesteps, we are implicitly performing this clustering operation to distill the pertinent linguistic content at the top of the stack. The reduction in timesteps is also favorable as regards learning attention, the rationale being that as the decoder examines all the frames of the encoder to extract attention parameters, it is useful to aggregate relevant information so that it has a smaller set to work with, which helps in speeding up the computation and in helping the model learn alignment. 

In order to reduce the number of input timesteps, we accumulate two neighboring frames, and then pass the concatenated features along to the bidirectional RNN layer above. In our experiments, we use a stack of $2$ recurrent reduction layers, resulting in an overall reduction in the number of timesteps by a factor of $4$. 

The basic unit of the hierarchical recurrent encoder is the bi-directional GRU. This bi-directional Gated Recurrent Unit (shown separately in the diagram as $L\to R$ and $R \to L$) passes over the input sequence twice: left to right and from right to left, and concatenates the two passes. Each GRU has 150 hidden units, and outputs a dimensionality 150xT, where T is input sequence length. After concatenation of the $L\to R$ and $R\to L$ outputs, one gets a dimensionality 300xT. The sequence length itself is reduced to T/2 after GRU1 and to T/4 after GRU2 as a result of accumulating 2 neighbouring frames at each step. GRU0 is a pre-processing recurrent unit that does not have this accumulation and reduction of time steps. The details of the pyramidal encoding, with tensor sizes after each step are shown in Figure \ref{fig:encoder}.

\subsection{Attention Decoder}
The decoder architecture is inspired by the Tacotron TTS setup \cite{Tacotron}. As in the Tacotron work, the decoder has the following components:
\begin{enumerate}
    \item Prenet 
    \item Attention RNN
    \item Decoder RNNs with residuality 
\end{enumerate}

We describe the components in more detail below. However, before doing so, it is useful to have in mind an overall picture of how the data flows through the decoder stack. To that end, we present a brief description of the calculations at a high level.

The decoder's task is to transform linguistic content from the source speaker to that of the target speaker in a temporal way, conditioned on frames generated previously. The linguistic content is provided by the hierarchical encoder described previously, which condenses the source speaker's utterances into an intermediate hidden representation embodying linguistic content. The decoder's task is therefore to ingest this linguistic content, and imbue it with the target speaker's voice characteristics. In the current setup, the DNN implicitly adds the target speaker's voice characteristics (i.e. duration, pitch) to the encoder summary. It is designed as a complex stack of unidirectional RNNs trained to emit output spectrogram frames conditioned on all previous frames emitted, together with the encoder's representation of the context.

The attention modeling ensures that the target's spectrogram frames are aligned with the appropriate frames of the input. Attention computations are ubiquitous in sequence to sequence modeling. While decoding output sequence frames - this could be in any general sequence modeling task, such as NMT, ASR, or TTS - attention helps to focus on the
appropriate frame of the input sequence so that the decoder is able to decide what it should emit in a more precise way. This aspect is especially important when the sequence length becomes large, for the decoder's task becomes much more difficult in emitting sequential output based on a single, global context that the encoder provides. Moreover, it is seen in experiments that attention modeling is essential for the system to generalize to unseen input. Our experiments seem to be in line with the notion that for the speech model to perform well on unseen data, it is in fact necessary for the model to learn proper alignment. 

We now proceed to describe in more detail the components of the decoder. 

\subsubsection{Prenet}
As with the encoder, we transform target data through a set of bottleneck layers (two in total) using dropout. We use dropout in order to regularize the model and prevent overfitting, and hence it is a very essential component. 
We use a stack of $2$ prenet layers (full connections with ReLU non-linearity and a dropout ratio of $0.5$) yielding vectors of size $256$ and $128$ respectively. 

\subsubsection{The Hybrid Content-Location Attention Model}
We use attention modeling \cite{Luong,Chorowski,Tacotron2} as a way of focusing the generator on the most relevant section of the input sequence. We have a state sequence output by the encoder (hidden) units at the top of the hierarchical stack: $h = (h_{1},\cdots{}h_{L})$. The sequence output by the decoder units is $s = (s_{1},\cdots{}s_{T})$. The input spectrogram sequence is  $x = (x_{1},\cdots{}x_{L})$ and the output spectrogram sequence is $y = (y_{1},\cdots{}y_{T})$. At the ith step of the generation process, the recurrent sequence generator, the RNN, generates state $s_{i}$ by using the $y_{i}$'s up to that point, the previous $s_{i-1}$ and the hidden encoder output $(h_{1},\cdots{}h_{L})$.
The attention model is used to inform the generator which encoder states $h_{j}$ are important for the generation of this $s_{i}$, and this is done with an attentional neural network, which learns to produce the attention or alignment vector $\alpha_{i}$, which is a vector of normalized importance weights used to weight the hidden encoder state $h$. This is then used to produce the context vector $c_i$, which is a weighted sum of the encoder states $ h_j$:

\begin{align}
s_i = RNN(s_{i-1}, [c_i, y_{i-1}]) \\
c_i = \sum_j{\alpha_{ij} h_j}
\end{align}

The context vector $c_i$, concatenated with the spectogram output prediction of the previous time step $y_{i-1}$ is used to condition the production of the decoder output for the current time step $s_{i}$.

The attention vector $\alpha_{ij}$ is obtained by softmax normalization (to between 0 and 1) with a $\beta$ temperature parameter, over the scores $e_{ij}$.

\begin{equation}
    \alpha_{ij} = softmax(\beta e_{ij})
\end{equation}
where $\beta$ is the softmax temperature that sharpens the attention (\cite{Chorowski}).

The scores or un-normalized attention energies $ e_{ij}$ is the central part of the attention modeling, and is done for each hidden encoder state $h_j$ separately.
There are two ways of calculating these attention energies.
\textbf{Content based attention} is dependent on the content or encoder hidden state:
$e_{ij} = score(s_{i-1},h_{j})$.
\textbf{Location based attention} is dependent on the location of the previous generator state, or where the attention was previously focused:
$e_{ij} = score(alpha_{i-1})$. This is normally implemented as a 1-D convolutional kernel (with learnt weights) centred around the previous position.
We use a hybrid attention model, with both content and location based scoring.
Location scoring is done by convolving the previous attention $\alpha_{i-1}$ with $F$. This is then combined with content scoring:
\begin{align} 
f_i = F*\alpha_{i-1} \\
e_{ij} = v^T\tanh((W_1 s_{i-1})^T (W_2 h_j) + U f_{ij}) \label{attn}
\end{align}
where vector $ v$ and matrices $F$, $W1$, $W2$ and $U$ are trainable weights, implemented as a feed-forward neural network. We use a form inspired by Luong's multiplicative attention mechanism \cite{Luong} to determine the mapping between hidden units and attention energies in equation \ref{attn}.

\begin{figure}
    \centering
    \includegraphics[width=0.90\linewidth]{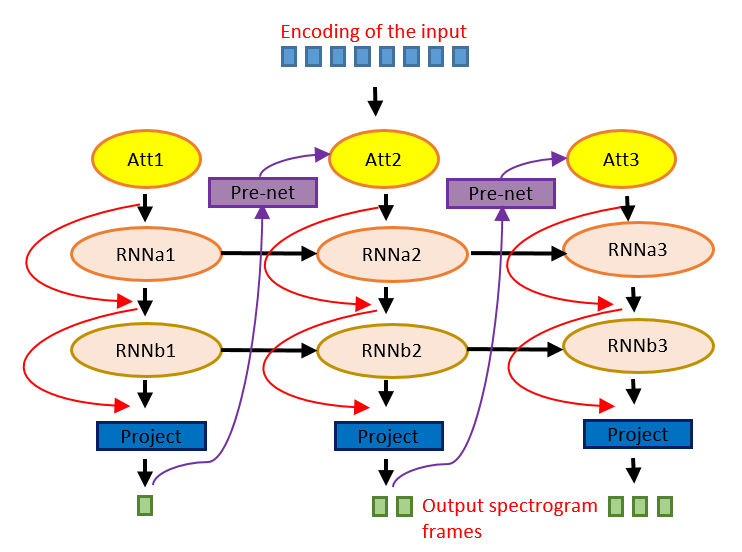}
    \caption{The decoder RNN. Att represents the Attention RNN, RNNa and b represent the first and second layers of the decoder RNNs. Red arrows indicate residual connections and purple arrows indicate the generated output being fed back to the attention RNN (along with input) to generate the attention output. Output of the second decoder is transformed to the dimensions of the output spectrogram using a fully connected layer (Project).}
    \label{fig:decoder}
\end{figure}

\subsection{Decoder RNNs with residuality} 
The attention RNN's output is now processed by two RNN layers with residuality, before transforming them back to audio frames.

\begin{align} 
g_i^1 = RNN^1(s_i, g_{i-1}^1) + s_i  \label{eqn:dec_residual1} \\
g_i^2 = RNN^2(g_i^1, g_{i-1}^2) + g_i^1 \label{eqn:dec_residual2}
\end{align}

This is depicted in the equations (\ref{eqn:dec_residual1}), (\ref{eqn:dec_residual2}). Here, the superscripts $1,2$ represent the first and second decoder layers. The second term in these equations contain the residual signal from the input. In this case, $s_i$ represents the output from the attention RNN and $g_i^1$ and $g_i^2$ denote the hidden units from the first and second decoder layers. We use the same number of dimensions ($300$) in all the decoder RNN layers. 

Finally, the output of the last residual decoder layer is transformed back to the dimensions of the output ($80$ bins) by sending it to a fully connected layer and adding a ReLU non-linearity to it. 

\section{Autoencoder pretraining and transfer learning} \label{section:AEPretraining}
Voice conversion with DNNs for parallel data is a difficult undertaking owing to the lack of availability of large multispeaker voice conversion datasets. To get around this problem, we first pretrain our network as an {\it autoencoder} with a large single speaker TTS corpus \cite{ljspeech17}, with the source and target voices being the same. After this network is trained - a guideline for this is to see if system learns alignment - we adapt the network for the smaller, multispeaker voice conversion data. 

Transfer learning can be seen as a way to mitigate data insufficiency problems in the speech domain. This is particularly trenchant owing to the lack of availability of good quality speech datasets (large corpuses, and with sufficient diversity) that can be obtained inexpensively.

The system is trained using the L1 loss between source and target voices. The Adam optimizer is used with a learning rate of $10^{-4}$ for the pretraining task, and $0.5\times 10^{-5}$ for the voice adaptation task. The optimizer parameters $\beta_1,\beta_2$ were $0.9$ and $0.999$ respectively. 

\section{Experimental setup}
Our experimental procedure consists of two steps, as mentioned in section \ref{section:AEPretraining}. We first pretrain the network with a large single-speaker corpus in which the source and the target are the same. After this, we allow the network to adapt to the desired source and target data. 

\subsection{Datasets}
For autoencoder pretraining, we use the LJSpeech dataset \cite{ljspeech17}. This dataset contains $13100$ short utterances from a single female speaker reading passages from $7$ audio books, with a total audio amounting to about $24$ hours recorded on a Macbook Pro in a home setting with a sampling rate of $22050\, \mathrm{Hz}$. 
The main task is to perform voice conversion (by adapting the pretrained network trained above) on the much smaller CMU Arctic dataset \cite{CMU-Arctic} containing $1132$  utterances from several speakers. We used the male speakers "bdl", "rms" and the female speakers "clb" and "slt" for experiments. The training/test/validation split was $1000$, $66$ and $66$ respectively. Since this corpus has a sampling rate of $16000\, \mathrm{Hz}$, we upsample this dataset to $22050\, \mathrm{Hz}$, generate audio through the pipeline and then downsample it back to the original sampling rate. This measure was adopted instead of downsampling the large corpus to the target sampling rate because we found that the system was unable to learn at the lower rate of $16\, \mathrm{kHz}$.
 
\begin{figure}
    \centering
    \includegraphics[width=0.99\linewidth]{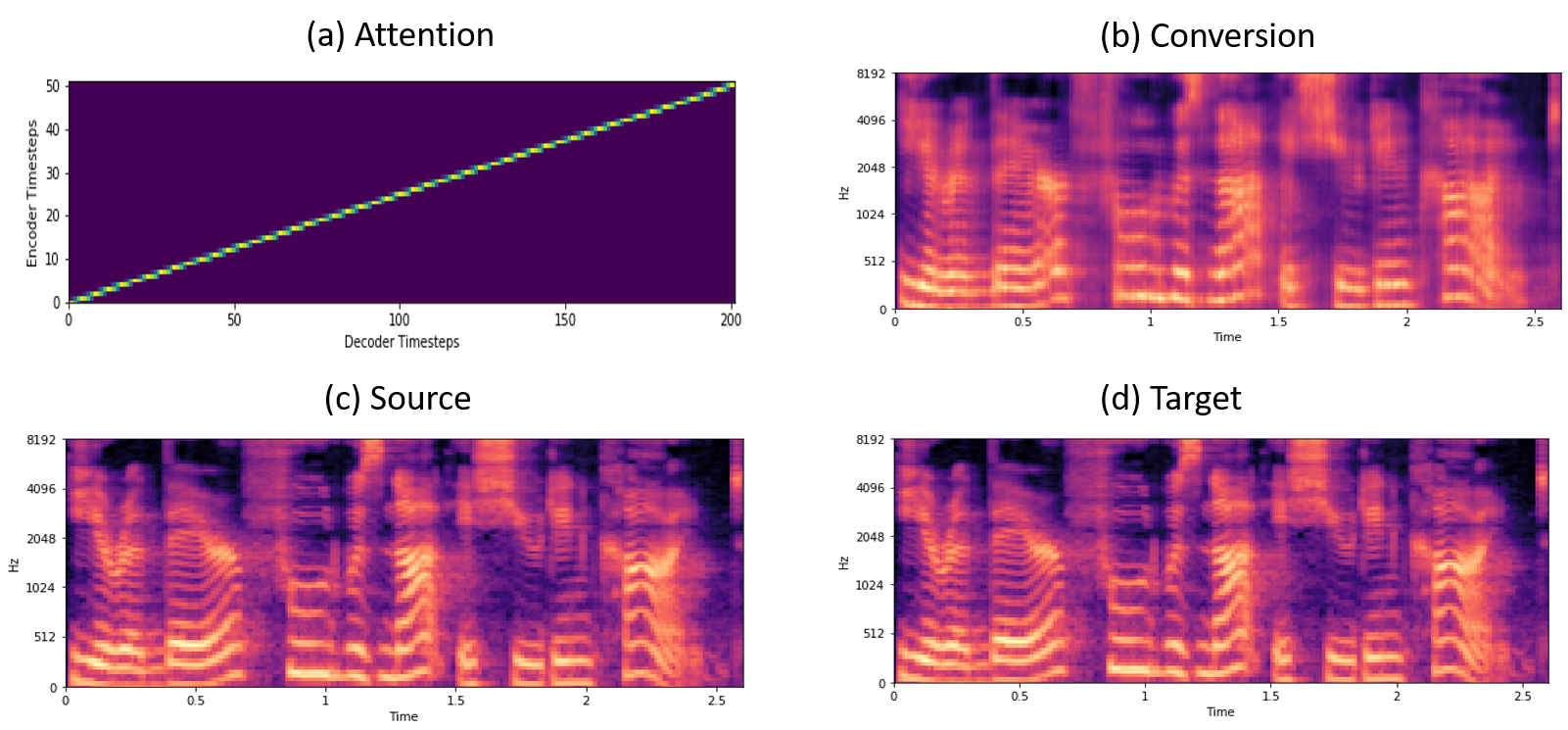}
    \caption{Feature extractor, depicted through attention alignment and mel spectrograms produced by training the network to produce ljspeech voices, with source and target being the same.}
    \label{fig:pretraining}
\end{figure}

\subsection{Example Conversions}
In figure \ref{fig:pretraining}, we present visualizations of source and target spectrograms, conversion and alignment curve for the pretrained autoencoder feature extractor using the large LJSpeech corpus. The alignment curve in this case shows more decoder timesteps than the encoder (by a factor of $4$) because of the hierarchical encoding scheme which reduces the number of timesteps in the encoder.

\begin{figure}
    \centering
    \includegraphics[width=0.99\linewidth]{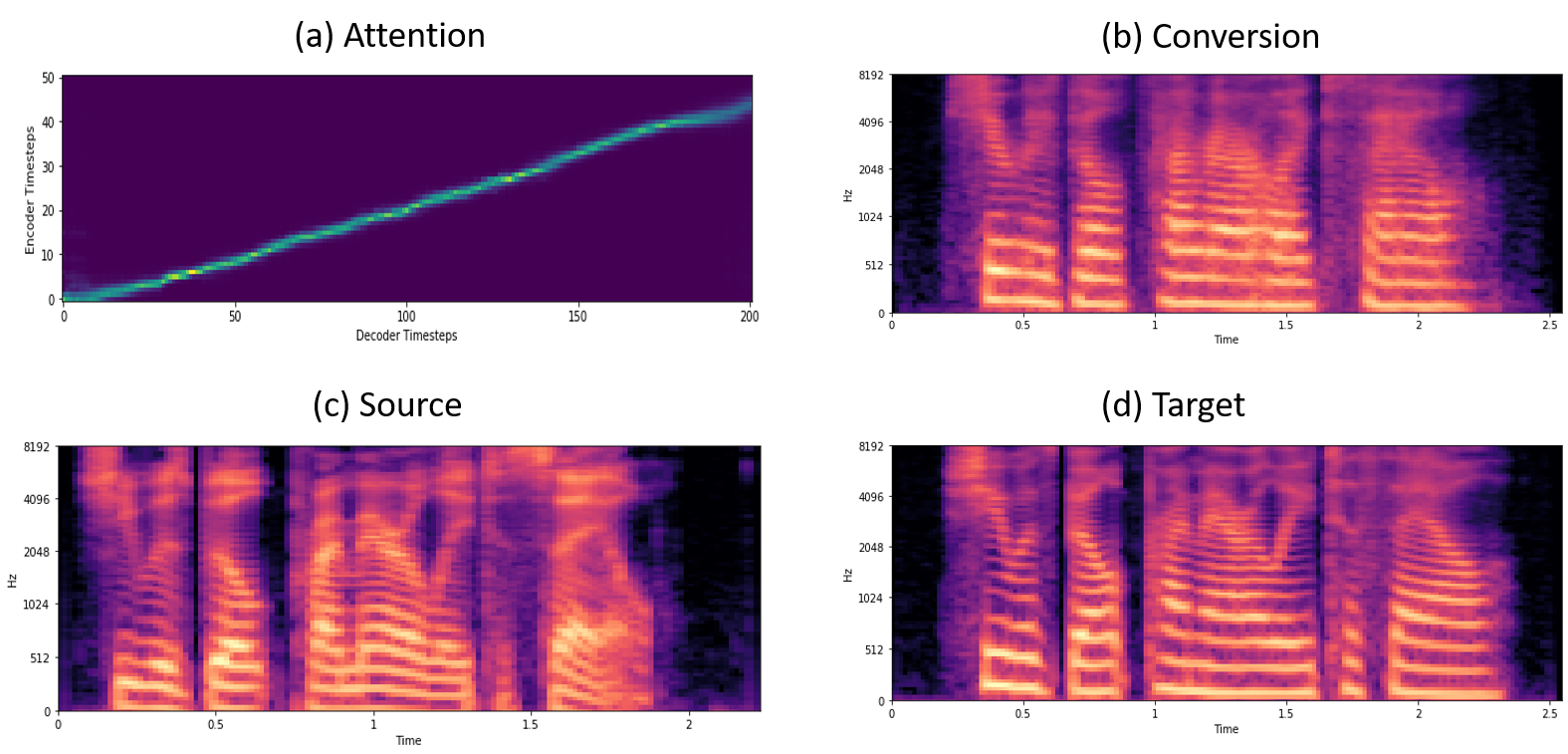}
    \caption{Voice conversion from male (bdl) to female (slt) voice, depicted through attention alignment and mel spectrograms produced by adapting to small CMU Arctic voice corpus.}
    \label{fig:conversion-adaptation}
\end{figure}


In figure \ref{fig:conversion-adaptation}, we present corresponding visualizations for the transfer learning experiment wherein we convert from male (bdl) to female (slt) voice. Starting with a network whose weights are pretrained with the large LJSpeech corpus as an autoencoder, we allow the network to adapt to the smaller CMU-Arctic dataset, using $1000$ paired training examples. As can be seen, while the conversion is plausible, the transfer learning spectrogram is somewhat `blurry' owing to the limited amount of data and the use of the L1 loss, which makes the spectrograms appear oversmoothed. While the alignment curve is more or less linear, it has a few 'kinks' (unlike the ljspeech curve) in keeping with the slight differences that arise in the alignment path as compared with the case where both source and target are the same.


\input{tables/ExperimentSettings.tex}



\subsection{Wavenet Implementation} 
We use a popular open source wavenet implementation \cite{r9y9-wavenet} available online to recover audio from mel spectrograms. Wavenet is an autoregressive architecture \cite{wavenet-original} especially designed for audio generation. Related architecutures have been used for generative modeling tasks in other domains: ByteNet \cite{ByteNet} for text, PixelCNN \cite{PixelCNN} for images and Video Pixel Net \cite{VideoPixelNet} for videos. This type of architecture, at a high level works on a temporal (in the sense that there is a certain temporal ordering of data) basis by stacking dilated convolutions with exponentially growing receptive field sizes (e.g. $2$, $4$, $8$, $16$). Masking is carried out so as to only allow information from the past. In wavenet, instead of masking, one simply uses all the inputs from the past for the operations as the data already has an implicit temporal order to it. The architecture also uses gating and skip connections to allow better information flow through the network stack. 

A drawback of this type of architecture is that while training is fast, inference is slow owing to the sample level autoregressive nature of the setup, in that every sample generated is conditioned on all previous samples; the upshot being that with raw audio ($22000$ samples per second), the calculations become extremely expensive. To alleviate these issues, themes from flow based generative modeling techniques (with some of the ideas originally proposed in order to improve the expressiveness of VAE priors \cite{NormalizingFlows, iaf} by successively transforming them) were adapted for fast inference during the sampling stage \cite{ParallelWavenet,WaveGlow}. 


To use as a vocoder backend, we present the wavenet with mel-spectrograms as conditioning features. These are upsampled (with transpose convolutions) to match the target rate with $4$ upsampling layers. This network has $24$ layers with $(512,512,256)$ channels for residuals, gating and skips respectively. The setup uses mixtures of logistics to model the $16$ bit ($65536$ bins) raw waveform. This architecture was also used to compare against the WaveGlow implementation in \cite{WaveGlow}. A more extensive list of hyperparameters is available online \cite{r9y9-wavenet}.




\section{Conclusions}
In this work, we demonstrated a way to overcome data limitations (an all too common malady in the speech world) with a trick to extract linguistic features by pretraining with a large corpus so that it learns to reconstruct the input voice. These features serve as a useful starting point for transfer learning in the limited data corpus. The architecture proposed is slightly elaborate, in that it resorts to hierarchically reducing the number of timesteps on the encoder side. The basis for this proposal was in keeping with the fact that the content embedded in the input waveforms - viewed as words or phoneme like entities - is much smaller than the size of the waveforms (5 words vs 100 audio frames, 10 phonemes vs 100 audio frames, etc.). With this intuition, the hierarchical reduction in timesteps is viewed as a mechanism to extract phoneme like entities by compressing the content in the input mel spectrogram. Our task is in a sense, to extract a style independent representation on the encoder side. The decoder then learns to inject the target speaker's content using exactly the same type of architecture as in the Tacotron works \cite{Tacotron, Tacotron2}. The output spectrograms are converted back to audio using a wavenet vocoder, yielding plausible conversions, demonstrating that our approach is indeed legitimate. 

The system is sensitive to hyperparameters. We noticed the capacity of the CBHG network is particularly important, and adding dropout at various places helps in generalizing to the small dataset. However, dropout also leads to 'blurriness'. Cleaning up the output is probably necessary with a postnet, which we have not implemented. 

We hope to release code and samples to allow for experimentation. 

\bibliography{VoiceConversion}
\bibliographystyle{IEEEtran}

\end{document}

%% file: tables/ExperimentSettings.tex
\begin{table}
\centering
\label{NetworkHyperparameters}
\caption{Network architecture hyperparameters. $Conv_k$-c-BN-ReLU-Dropout(f) denotes a convolution of width $k$, $c$ output channels, BatchNorm, ReLU, with a dropout of $f$(=$0.1$). $Conv_3$-$80$-Dropout(f)-ReLU-Linear denotes a convolution of width $3$, with $80$ output channels, dropout of $f$(=$0.1$), followed by ReLU and a linear projection to the same size output. Prenet layers are full connections (e.g. FC-$256$ would be a linear connection to an output of size $256$) but with dropout of $0.5$. All other network components use a dropout of $0.1$}
\begin{tabular}{|l|l|}
\hline
\multirow{1}{*}Encoder Prenet & FC-80-ReLU-Dropout(0.5) \\ 
CBH & $Conv_k$-BN-480-ReLU-Dropout(0.1)\\
    & $k=1,3,5,7,\cdots,25$ \\
    & Maxpool (stride=1) and stack\\
    & $Conv_3$-$80$-Dropout(0.1)-ReLU-Linear  \\
    & Highway layer stack of $4$ \\
BiGRU0 & 300 cells (f+b); Dropout(0.1)\\
Hierarchical BiGRU & 2 layers, 300 cells (f+b);\\ 
                   & Dropout(0.1) \\
                   \hline
 \multirow{1}{*}Decoder Prenet & FC-256-ReLU-Dropout(0.5) \\  
 & FC-256-ReLU-Dropout(0.5) \\
 Attention GRU & 600 cells; Dropout(0.1) \\
 Residual GRU 1,2 & 600 cells; Dropout(0.1) \\
 \hline
 \end{tabular}
\end{table}